\documentclass[reprint,superscriptaddress,onecolumn,longbibliography,amsmath,amssymb,aps]{revtex4-2} 
\usepackage{graphicx}
\usepackage{dcolumn}
\usepackage{bm}
\usepackage{hyperref}

\newcommand{\argmin}{\mathop{\rm arg~min}\limits}

\begin{document}

\title{Collaborative filtering based on nonnegative/binary matrix factorization} 
\author{Yukino Terui}
\affiliation{Department of Computer Science, Ochanomizu University, Tokyo, Japan}
\author{Yuka Inoue}
\affiliation{Department of Computer Science, Ochanomizu University, Tokyo, Japan}
\author{Yohei Hamakawa}
\affiliation{Corporate Research and Development Center, Toshiba Corporation, Kawasaki, Japan}
\author{Kosuke Tatsumura}
\affiliation{Corporate Research and Development Center, Toshiba Corporation, Kawasaki, Japan}
\author{Kazue Kudo}
\affiliation{Department of Computer Science, Ochanomizu University, Tokyo, Japan}
\affiliation{Graduate School of Information Sciences, Tohoku University, Sendai, Japan }

\begin{abstract}
Collaborative filtering generates recommendations by exploiting user–item similarities based on rating data, which often contains numerous unrated items.
To predict scores for unrated items, matrix factorization techniques such as nonnegative matrix factorization (NMF) are often employed. 
Nonnegative/binary matrix factorization (NBMF), which is an extension of NMF, approximates a nonnegative matrix as the product of nonnegative and binary matrices. 
While previous studies have applied NBMF primarily to dense data such as images, this paper proposes a modified NBMF algorithm tailored for collaborative filtering with sparse data. 
In the modified method, unrated entries in the rating matrix are masked, enhancing prediction accuracy. 
Furthermore, utilizing a low-latency Ising machine in NBMF is advantageous in terms of the computation time, making the proposed method beneficial.
\end{abstract}

\maketitle

\section{Introduction}

Collaborative filtering is often applied in recommendation systems that primarily serve Internet services, such as e-commerce and video distribution platforms~\citep{Herlocker2000,Su2009}.
The essence of collaborative filtering lies in generating personalized recommendations based on the intrinsic similarities between users and items. 
Collaborative filtering relies on training data, in which users assign scores or ratings to various items.
As it is common for users to omit ratings of specific items, leading to missing data, the central objective of collaborative filtering is to predict the scores for unrated items.
Matrix factorization techniques, particularly nonnegative matrix factorization (NMF)~\citep{Lee1999}, are frequently employed.
When using NMF for collaborative filtering, the ranking matrix $V$, whose entries are nonnegative, is approximated as the product of two nonnegative matrices $W$ and $H$, that is, $V\approx WH$.
The standard approach involves minimizing the difference between $V$ and $WH$.
In the optimization procedure, each element of $W$ and $H$ is constrained to be nonnegative.
While the multiplicative update algorithm is the most prevalent approach for NMF~\citep{Lee2000}, we focus on an alternative technique known as the nonnegative least-squares approach using the projected gradient method (PGM)~\citep{Lin2007}.
The convergence of the alternative update method for NMF was proved by \cite{Lin2007}.
Such an alternative optimization method is essential for solving nonnegative/binary matrix factorization (NBMF), which is an extension of NMF. 
\cite{OMalley2018} and \cite{Golden2021} used D-Wave's quantum annealers to solve quadratic binary optimization problems involved in NBMF and demonstrated a speedup compared with two classical solvers.

In recent years, Ising machines, initially designed to solve combinatorial optimization problems efficiently, have found new applications in the field of machine learning, expanding their scope beyond their original purpose~\citep{Kitai2020,Willsch2020,Nath2021}.
Ising machines are special-purpose computers for solving combinatorial optimization problems.
They are realized by several types of devices, such as quantum annealers ~\citep{Johnson2011}, 
digital processors based on simulated annealing~\citep{Yamaoka2016,Aramon2019,Yamamoto2020}, 
digital processors based on simulated bifurcation (SB)~\citep{Goto2019,Hidaka2023},
and coherent Ising machines~\citep{Inagaki2016,McMahon2016,Pierangeli2019}.
As Ising machines usually accept problems described by the Ising model or quadratic unconstrained binary optimization formulation, their application to machine learning requires hybrid methods that utilize both an Ising machine and a general-purpose computer (e.g., a CPU).
In NBMF, the matrix elements of $H$ are binary, whereas those of $W$ are real and nonnegative.
Therefore, an Ising machine is employed to accelerate the update of matrix $H$, whereas a general-purpose computer handles the update of matrix $W$. 
As the updates of matrices $H$ and $W$ are repeated alternately, NBMF inevitably involves a computation time overhead owing to the communication between the Ising machine and the CPU. 
The advantages and disadvantages of NMF and NBMF remain unclear in terms of solution quality, computation time, and applicability to sparse problems.

In this paper, we propose a novel approach for applying NBMF to collaborative filtering and demonstrate the advantages of utilizing a low-latency Ising machine to execute the proposed method.
Previous studies have employed NBMF for image analysis that deals with dense data matrices, where the majority of matrix elements have nonzero values~\citep{OMalley2018,Asaoka2020,Asaoka2023}.
By contrast, collaborative filtering involves sparse data matrices, with most elements remaining undetermined.
We propose a modified NBMF algorithm that masks undetermined elements within the data matrix to improve the prediction accuracy.
In addition, we compare NBMF with NMF in terms of solution quality and computation time, and investigate the dependency of these characteristics on the sparsity and size of the problem. 
To accelerate the NBMF algorithm, we used an SB-based machine implemented with a field-programmable gate array (FPGA)~\citep{Goto2021,Hidaka2023} that supports up to 2,048 spins and has full spin-to-spin connectivity (no need for minor embedding techniques required for local-connectivity Ising machines). 
Incorporating an SB-based machine to update the binary matrix elements yields a substantial reduction in the overall computational time required for NBMF compared with NMF. 

Furthermore, the low-latency characteristic of the SB-based machine is advantageous for executing the iterative method using a general-purpose computer (a CPU) and an Ising machine, alternatively, reducing the communication time between them.
It is also possible to use a cloud-hosted Ising machine for executing the proposed method. 
While a high-performance cloud-hosted Ising machine can significantly reduce computation time, the communication costs of accessing it may negate the benefits. 
Therefore, utilizing the low-latency system is crucial.
This study presents the first empirical evidence that NBMF, when implemented with a low-latency Ising machine, surpasses NMF in terms of both solution quality and overall computational efficiency.

\section{Problem formulation}

NBMF and NMF decompose a nonnegative $n \times m$ matrix $V$ into an $n \times k$ matrix $W$ and a $k \times m$ matrix $H$: 
\begin{equation}   
 V \approx WH,
  \label{eq:V}
\end{equation}
where $W$ is a nonnegative real matrix.
While $H$ is a binary matrix in which each element is $0$ or $1$ for NBMF, it is a nonnegative real matrix for NMF.
We assume that $n>k$ and $m>k$, which implies that NBMF and NMF provide low-rank matrix approximations of $V$.
The rank constraint is helpful to prevent overfitting.
Moreover, NBMF can be more resilient to overfitting due to the binary nature of matrix $H$.

In the context of collaborative filtering, $V$ is a rating matrix, where the $(i,j)$ element $v_{ij}$ represents user $i$'s rating of item $j$.
In matrix $W$, each row corresponds to a user, while each column represents a basis vector associated with user preferences.
In other words, $W$ consists of $k$ basis vectors, with each dimension being $n$.
Meanwhile, each column of $H$ represents the coefficient vector related to the corresponding item.
In NBMF, this coefficient vector indicates the combination of the selected basis vectors for the corresponding item.
In general, the rating matrix contains numerous unrated entries.
Matrix factorization techniques optimize $W$ and $H$ so that each rated entry in $V$ is well approximated by the corresponding element of $WH$.
Then, each unrated entry in $V$ is estimated by the corresponding element of $WH$.

The comparison between NMF and other collaborative filtering techniques has already been extensively studied~\citep{Lee2014,Singh2024}.
Compared to user-based and item-based collaborative filtering techniques, matrix factorization techniques demonstrated better performance in recommendation systems on multi-criteria datasets~\citep{Singh2024}.
In particular, NMF is scalable to large datasets and can capture individual user preferences. 
However, there has been no direct comparison between NMF and NBMF.
This paper focuses on comparing the two methods.

\section{Methods}

\subsection{Algorithm}

The approach to conducting matrix factorization involves minimizing $\|V-WH\|_F$, where the Frobenius norm is defined as $\| A \|_F=\sqrt{\sum_{i,j}A_{ij}^2}$, and $A_{ij}$ is the $(i,j)$ element of $A$. 
To achieve minimization, NBMF employs an iterative alternative update procedure as follows:
\begin{align}
 W &= \argmin_{X \in \mathbb{R}_{+}^{n \times k}} \left ( 
\| V - XH \|_F^2
 + \lambda_1 \| X \|_F^2
 \right),
  \label{eq:W} \\
 H &=  \argmin_{X \in { \{ 0,1 \} }^{k \times m}} \left(
\| V - WX \|_F^2
 + \lambda_2 \| X \|_F^2
 \right),
\label{eq:H} 
\end{align}
where $X$ is a matrix that corresponds to $W$ and $H$ in Eqs.~\eqref{eq:W} and ~\eqref{eq:H}, respectively.
Hyperparameters $\lambda_1$ and $\lambda_2$ are positive.

Matrix $W$ is updated row-by-row.
The objective function for the row vector $\bm{x}^\top$ of $W$ is given by
\begin{align}
f_{W}(\bm{x}) = \frac12\| \bm{v} - H^\top \bm{x} \|^2 
+ \frac{\lambda_1}{2} \| \bm{x} \|^2,
\label{eq:f_W}     
\end{align}
where $\bm{v}^\top$ is the corresponding row vector in matrix $V$.
We applied the PGM to minimize the objective function for each row vector, as detailed in Section~\ref{sec:PGM}.
The PGM was executed using a general-purpose computer.
In contrast, matrix $H$ is updated column-by-column.
The objective function for optimizing the column vector $\bm{q}$ ($\in\{0,1\}^k$) of $H$ is given by
\begin{align}
f_{H}(\bm{q}) = \frac12\| \bm{u} - W \bm{q} \|^2 
+ \frac{\lambda_2}{2} \| \bm{q} \|^2,
\label{eq:f_H}     
\end{align}
where $\bm{u}$ is the corresponding column vector in matrix $V$.
To minimize the objective function for each column, we employed an SB-based Ising machine, as Eq.~\eqref{eq:f_H} can be reformulated in the Ising model form (see to Section~\ref{sec:Ising} for details).

\begin{figure}[htb]
\centering
\includegraphics[width=0.25\linewidth]{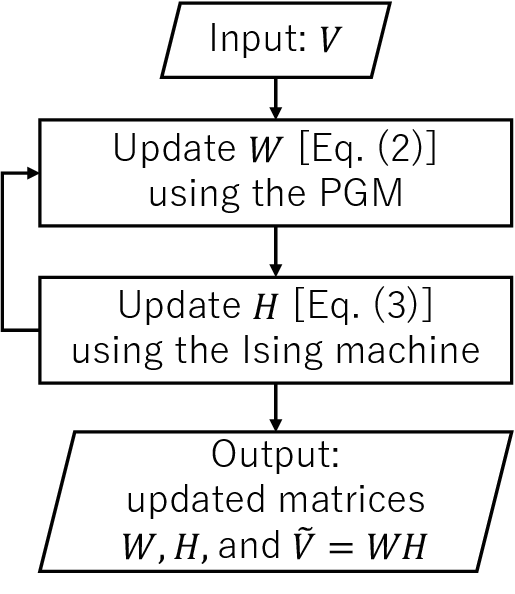}
\caption{Overall flow of NBMF. Matrix $\tilde{V}$ is an approximation of matrix $V$.}
\label{fig:chart}
\end{figure}

The overall flow of NBMF is illustrated in Figure~\ref{fig:chart}.
The process of updating matrix $W$, followed by the update of matrix $H$, was repeated for 10 iterations in this paper.

In this study, we compared NBMF with NMF. 
In NMF, Eqs.~\eqref{eq:W} and \eqref{eq:H} are also used; however, $X \in { \{ 0,1 \} }^{k \times m}$ in Eq.~\eqref{eq:H}  is substituted by $X \in \mathbb{R}_{+}^{k \times m}$.
Furthermore, each column vector $\bm{q}$ in Eq.~\eqref{eq:f_H} is nonnegative.
Equations \eqref{eq:f_W} and \eqref{eq:f_H}  were minimized using the PGM in NMF, and the computation was executed on a general-purpose processor (a CPU).

\subsection{\label{sec:PGM}Projected gradient method}

The PGM~\citep{Lin2007} for updating matrix $W$ minimizes Eq.~\eqref{eq:f_W}, and the gradient is given by 
\begin{align}
    \nabla f_{W}
  = -H (\bm{v} - H^\top \bm{x})
  + \lambda_1 \bm{x}.
  \label{eq:grad_W}
\end{align}
The update rule for $\bm{x}$ is given by
\begin{align}
 \bm{x}^{t+1}
  = P[\bm{x}^t - \gamma_t \nabla f_W (\bm{x}^t)],
  \label{eq:update}
\end{align}
where the projection is defined as
\begin{align}
 P[x_i] = 
  \begin{cases}
   0 & (x_i \leq 0), \\
    x_i & (0 < x_i < x_{\rm{max}}), \\
   x_{\rm{max}} & (x_{\rm{max}} \leq x_i).
  \end{cases}
\end{align} 
In this study, we set $x_{\rm{max}}=1$ as the upper bound of $x_i$. 
The learning rate $\gamma_t$ was adjusted at each step $t$ to satisfy the following inequality:
\begin{align}
    f_W(\bm{x}^{t+1})-f_W(\bm{x}^t)\le \sigma\nabla f_W(\bm{x}^t)^\top(\bm{x}^{t+1}-\bm{x}^t),
    \label{eq:pgm_ineq}
\end{align}
where $\sigma=0.01$ in our experiments.
Initially, we assigned $\gamma_{t-1}$ to $\gamma_t$ ($\gamma_0=1$).
If $\gamma_t$ satisfies Eq.~\eqref{eq:pgm_ineq}, it is repeatedly divided by $\beta$, where we set $\beta=0.1$ in our experiments, while the inequality holds.
If $\gamma_t$ does not satisfy Eq.~\eqref{eq:pgm_ineq}, it is repeatedly multiplied by $\beta$ until the inequality is satisfied.
Following this adjustment, we calculated $\bm{x}^{t+1}$ using Eq.~\eqref{eq:update}.
This procedure is repeated until $\|\bm{x}^{t+1}-\bm{x}^t\|\ll\epsilon$, where $\epsilon=10^{-7}$ in our experiments.

\subsection{Masking procedure}

Given that the rating matrix is typically sparse, the handling of unrated entries has a significant impact on the performance of collaborative filtering.
A straightforward approach is to assign a rating of zero to unrated entries, which is a simple and practical choice. 
Another method for handling unrated entries is to introduce a mask matrix of the same size as matrix $V$ after assigning them a zero rating. 
The elements of the mask matrix $M$ are defined as follows:
\begin{align}
    M_{ij} =
    \begin{cases}
        1 & (V_{ij}\neq 0), \\ 0 &  (V_{ij}= 0).
    \end{cases}
\end{align}
For collaborative filtering, we propose a modified NBMF method in which the masked matrix is decomposed as
\begin{align}
    M\circ V \approx M\circ(WH),
\end{align}
where $\circ$ denotes the Hadamard product $(M\circ V)_{ij}=M_{ij}V_{ij}$.

In the modified NBMF algorithm, the objective function for updating matrix $W$, as defined by Eq.~\eqref{eq:f_W}, is replaced with
\begin{align}
    f_{W}(\bm{x}) = \frac12\| \tilde{\bm v} - \tilde{H}^\top \bm{x} \|^2 
    + \frac{\lambda_1}{2} \| \bm{x} \|^2.
\label{eq:f_W.2}     
\end{align}
When updating the $i$th row, the $j$th element consists of $\tilde{v}_j=M_{ij}V_{ij}$ and $(\tilde{H}^\top\bm{x})_j=\sum_l M_{ij}H_{lj}x_l$.
Similarly, the objective function for updating matrix $H$, as expressed in Eq.~\eqref{eq:f_H}, is replaced by
\begin{align}
    f_{H}(\bm{q}) = \frac12\| \tilde{\bm u} - \tilde{W} \bm{q} \|^2 
    + \frac{\lambda_2}{2} \| \bm{q} \|^2.
\label{eq:f_H.2}     
\end{align}
When updating the $j$th column, the $i$th element consists of $\tilde{u}_i=M_{ij}V_{ij}$ and $(\tilde{W} \bm{q})_i=\sum_l M_{ij}W_{il}q_l$.

\subsection{Ising formulation}
\label{sec:Ising}

The Ising machine (the SB-based machine in this study) seeks spin configurations that minimize the energy of the Ising model defined by
\begin{align}
 E = -\frac12\sum_{i,j} J_{ij} s_i s_j + \sum_{i} h_i s_j.
 \label{eq:E}
\end{align}
Here, $s_i=\pm 1$ represents the $i$th spin, $J_{ij}$ is the coupling coefficient between the $i$th and $j$th spins, and $h_i$ is the local field on the $i$th spin.
For minimizing Eq.~\eqref{eq:f_H}, $J_{ij}$ and $h_i$ are given as follows:
\begin{align}
    J_{ij} &=
    \begin{cases}
        -\frac12\sum_rW_{ri}W_{rj} & (i\neq j), \\
        0 & (i=j),
    \end{cases} 
    \label{eq:Jij}\\
    h_i &=\frac12\left( \sum_r W_{ri}\left(\sum_j W_{rj}-2u_r\right)
    +\lambda_2\right). 
    \label{eq:hi}
\end{align}
For minimizing Eq.~\eqref{eq:f_H.2} to update the $j$th row of $M\circ (WH)$, $W_{rl}$ in Eqs.~\eqref{eq:Jij} and \eqref{eq:hi} are replaced with  $(\tilde{W})_{rl}=M_{rj}W_{rl}$.

\subsection{Simulated-bifurcation-based Ising machine}

The SB algorithm, which is based on the adiabatic evolution in classical nonlinear systems that exhibit bifurcation, was introduced to accelerate combinatorial optimization~\citep{Goto2019,Tatsumura2020,Goto2021}. The SB algorithm has several variants, including adiabatic, ballistic, and discrete SB.
In this study, we employed the ballistic SB method, whose update rule is described below~\citep{Goto2021}:
\begin{align}
 y_i(t_{k+1}) &= y_i(t_k) + \left\{ -[a_0-a(t_k)]x_i(t_k) -\eta h_i
+ c_0\sum_{j}J_{ij}x_j(t_k)\right\}\Delta_t,
\label{eq:y.bSB}\\
x_i(t_{k+1}) &= x_i(t_k) + a_0 y_i(t_{k+1})\Delta_t,
\label{eq:x.bSB}
\end{align}
where $x_i$ and $y_i$ are real numbers corresponding to the $i$th spin; $a_0$, $c_0$, and $\eta$ are positive constants; and $a(t)$ is a control parameter that increases from zero to $a_0$.
The time increment is $\Delta_t$; thus, $t_{k+1}=t_k+\Delta_t$. 
After updating $x_i$ at each time step, if $|x_i|>1$, we replace $x_i$ with $\mathrm{sgn}(x_i)=\pm 1$ and set $y_i=0$.

\begin{figure}[htb]
\centering
\includegraphics[width=0.7\linewidth]{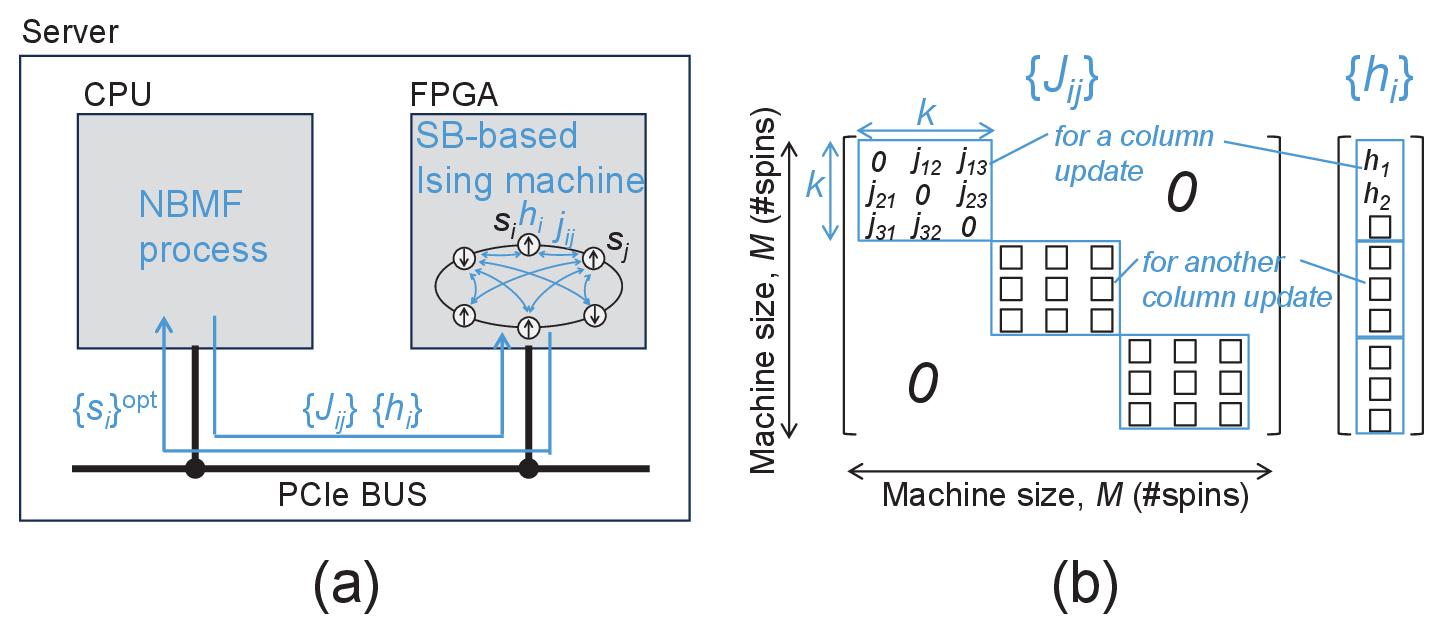}
\caption{
SB-based Ising machine. (a) System configuration. (b) Packing of multiple small Ising problems as a large Ising problem.}
\label{fig:SBM}
\end{figure}

In our experiments, we employed a device with the FPGA implementation of the SB algorithm (SB-based Ising machine) to minimize Eqs.~\eqref{eq:f_H} and ~\eqref{eq:f_H.2}.
The SB-based Ising machine (Figure~\ref{fig:SBM}) can solve fully-connected 2,048-spin Ising problems (the machine size $M$ is 2,048), featuring a computational precision of 32-bit floating points and a system clock frequency of 259 MHz. 
As shown in Figure~\ref{fig:SBM}a, the FPGA (Intel Stratix 10 SX 2800 FPGA) on the board (Intel FPGA PAC D5005 accelerator card) is connected to a CPU (Intel Core i9-9900K, 3.60 GHz) via a PCIe bus (Gen 3$\times$16, peak bandwidth of 15.75 GB/s). 
The NBMF process is executed using a CPU; however, the Ising problems described in Eqs.~\eqref{eq:Jij} and \eqref{eq:hi} are transferred/solved (offloaded) to/using the SB-based Ising machine. 
The computation times shown in Figures~\ref{fig:base}, \ref{fig:mask}, and \ref{fig:time} include the processing times of the CPU and FPGA and the data transfer times (overhead times) between them. 
The NMF process was executed only on the CPU (no data transfer time). 
The column update problems involved in updating matrix $H$  [Eq.~\eqref{eq:f_H}], each formulated as an Ising problem of size $k$ [Eqs.~\eqref{eq:Jij} and \eqref{eq:hi}], are independent and thus can be processed simultaneously.  
By packing the multiple-column update problems as a large Ising problem, as shown in Figure~\ref{fig:SBM}b  [placing the small $J$ matrices on the diagonal line with zero padding to the remaining off-diagonal components], we solve $\lfloor M/k \rfloor$ column update problems simultaneously using the SB-based Ising machine with size $M$, where $\lfloor A \rfloor$ is the floor function of $A \in \mathbb{R}$.

\subsection{Data preparation}

In this study, we used the MovieLens 1M dataset~\citep{Movielens}; 
Netflix Prize data\footnote{Netflix Prize data, https://www.kaggle.com/datasets/netflix-inc/netflix-prize-data}; Yahoo! Music user ratings of songs with artist, album, and genre meta information\footnote{Yahoo! Music User Ratings of Songs with Artist, Album, and Genre Meta Information, v.~1.0,
http://webscope.sandbox.yahoo.com/}; and the CiaoDVD dataset~\citep{Guo2014}.
These datasets were sparse, as shown in Table~\ref{tab:data}.
The numbers of users and items presented in Table~\ref{tab:data} are the dataset sizes imported for the calculation in this study.
The filling rate, which is the proportion of rated entries, differs among the datasets.
To compare the results of these datasets, we extracted data from them to create a rating matrix with a specified filling rate.

\begin{table}[htb]
 \centering
 \caption{Dataset sizes (the numbers of users and items) and filling rates used in this study. 
The original sizes of the Netflix and Yahoo datasets are significantly larger: the Netflix dataset includes 480,189 users, and the Yahoo dataset includes 1.8 million users.}
 \label{tab:data}
    \begin{tabular}{|l|r|r|r|}
        \hline
        Dataset &  Users &  Items & Filling rate\\
        \hline
        MovieLens & 6,040 & 3,706 & 4.47\%\\
        Netflix & 432,229 & 1,406 & 1.15\%\\
        Yahoo & 2,677 & 126,478 & 0.30\%\\
        CiaoDVD & 21,019 & 71,633 & 0.11\%\\
        \hline
    \end{tabular}
\end{table}

The method for extracting data at a specified filling rate is as follows. 
First, we sorted the columns in descending order by the percentage of filled elements in each column and then sorted the rows similarly.
Next, we selected an $n\times m$ matrix whose $(1,1)$ element coincides with the first-row and first-column element of the sorted table, and calculated the filling rate of the matrix.
By shifting the $(1,1)$-element location by one row and one column in the sorted table, we repeated the calculation of the filling rate.
The $n\times m$ matrix with the closest filling rate to the desired filling rate was selected as the rating matrix.

\subsection{Parameter settings and evaluation}

By extracting data from each dataset, we constructed a rating matrix in which 20\% of the elements were rated unless otherwise specified.
The numbers of users (rows) and items (columns) in the rating matrix are $n=250$ and $m=500$, respectively, and the number of features is set to half the number of users, that is, $k=n/2$, unless otherwise specified.
For the learning process, which involved the execution of NBMF/NMF, we concealed 20\% of the rated elements together with the unrated ones.
To evaluate the performance, we used the root mean squared error (RMSE) of the rated elements:
\begin{align}
    \sqrt{\frac{1}{|\mathcal{D}|}\sum_{(i,j)\in\mathcal{D}}(v_{ij}-r_{ij})^2},
    \label{eq:rmse}
\end{align}
where $\mathcal{D}$ is the set of rated elements, and $|\mathcal{D}|$ is the number of rated elements.
$v_{ij}$ is user $i$'s rating for item $j$, and $r_{ij}$ denotes the corresponding predicted rating.

We set the hyperparameters in Eqs.~\eqref{eq:W} and \eqref{eq:H} as $\lambda_1=10^{-2}$ and $\lambda_2=10^{-5}$, which were tuned for the MovieLens dataset using a grid search. 
Parameters in Eqs.~\eqref{eq:pgm_ineq} and \eqref{eq:y.bSB} were also tuned for the case of the MovieLens dataset.
Although the optimal values may depend on the dataset and matrix size, we used the ﬁxed values for simplicity.

In our experiments, we applied NBMF and NMF to the same rating matrix.
To ensure a comprehensive evaluation, we divided the rated elements into five distinct sets and performed five trials, masking one set at a time.
The average was calculated for five trials unless otherwise specified.

\section{Results and discussion}

\begin{figure}[htbp]
\centering
\includegraphics[width=0.4\linewidth]{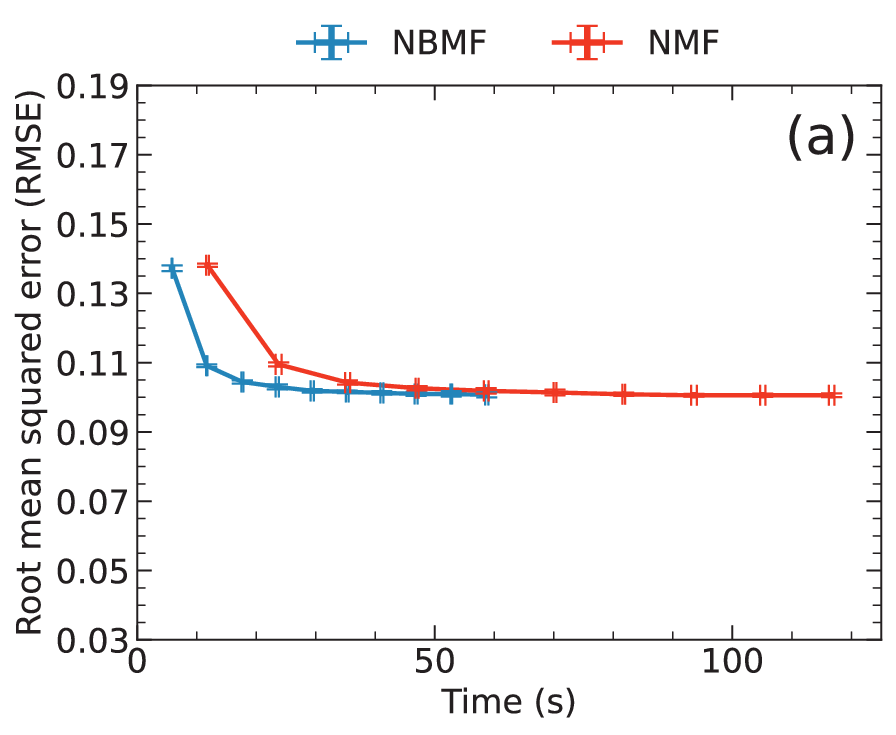}
\includegraphics[width=0.4\linewidth]{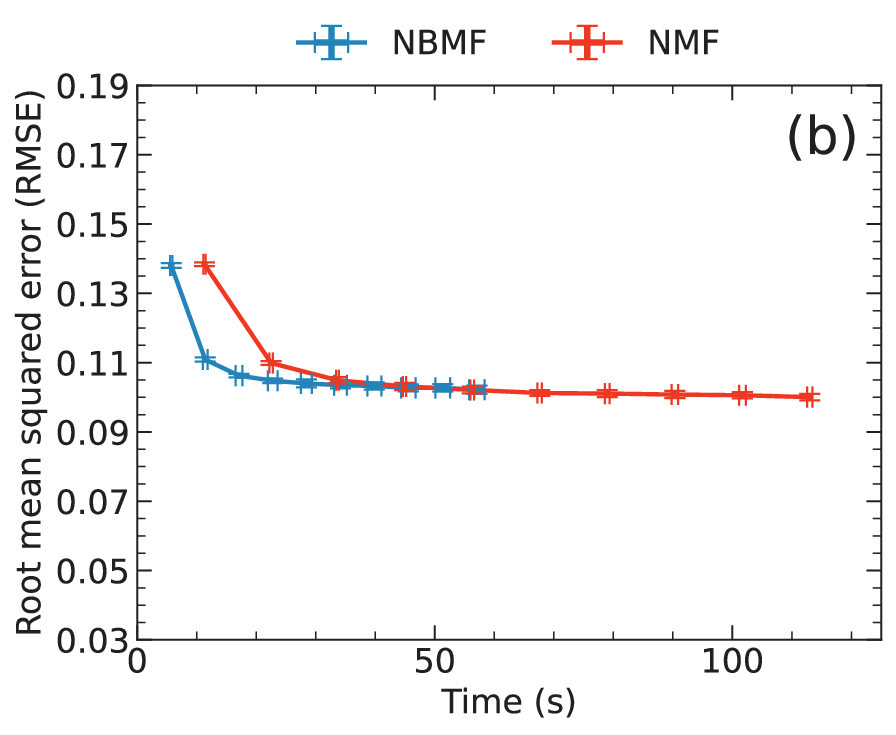}
\includegraphics[width=0.4\linewidth]{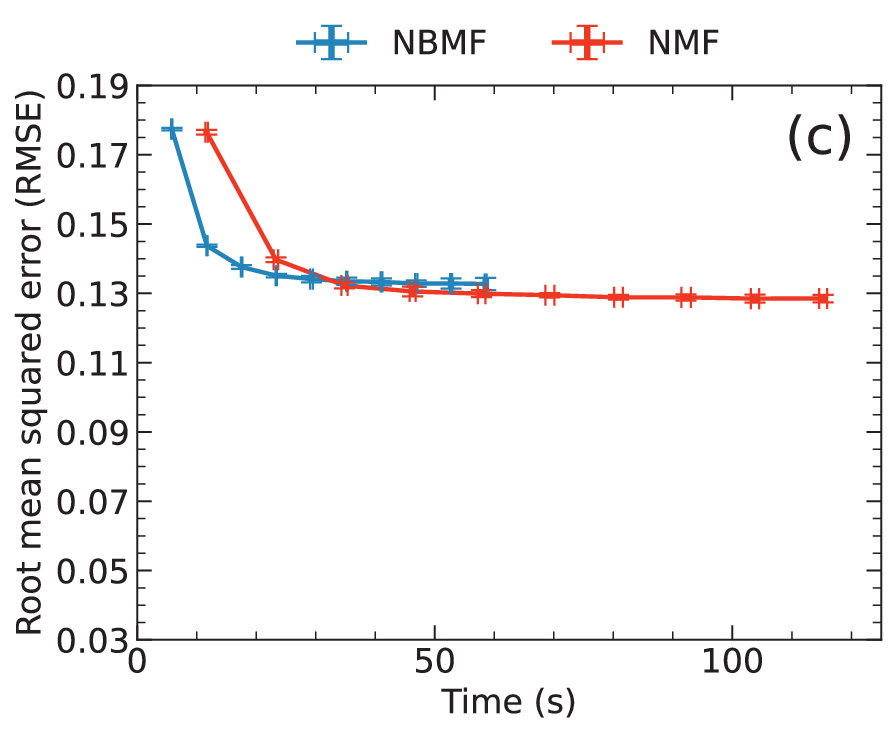}
\includegraphics[width=0.4\linewidth]{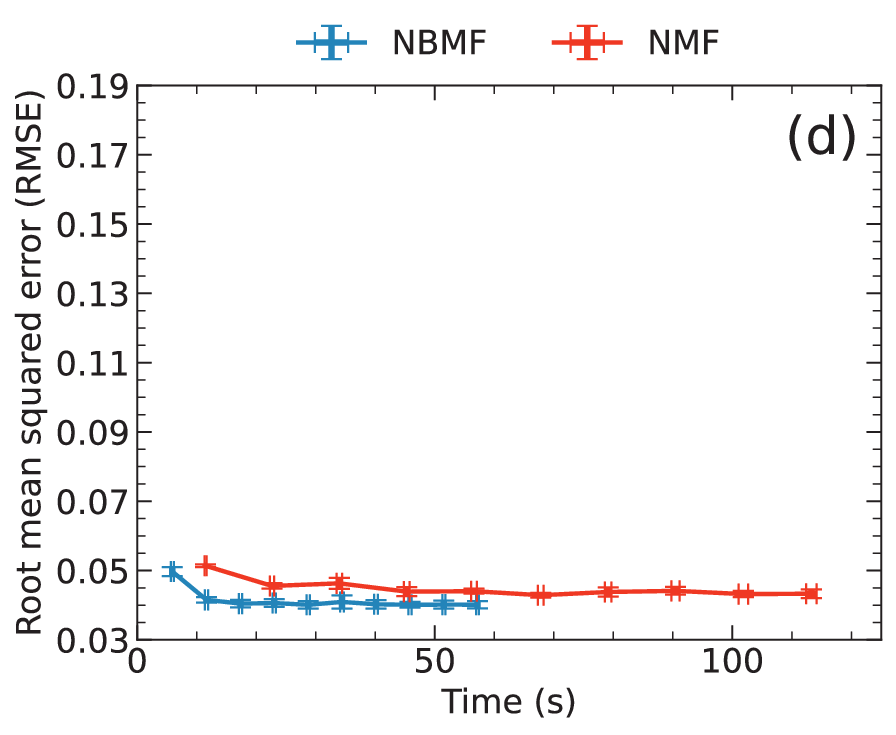}
\caption{RMSE and computation time at each epoch, averaged over five trials, 
for (a) MovieLens, (b) Netflix, (c) Yahoo, and (d) CiaoDVD datasets.
The error bars denote the standard deviation.}
\label{fig:base}
\end{figure}

\begin{figure}[htbp]
\centering
\includegraphics[width=0.4\linewidth]{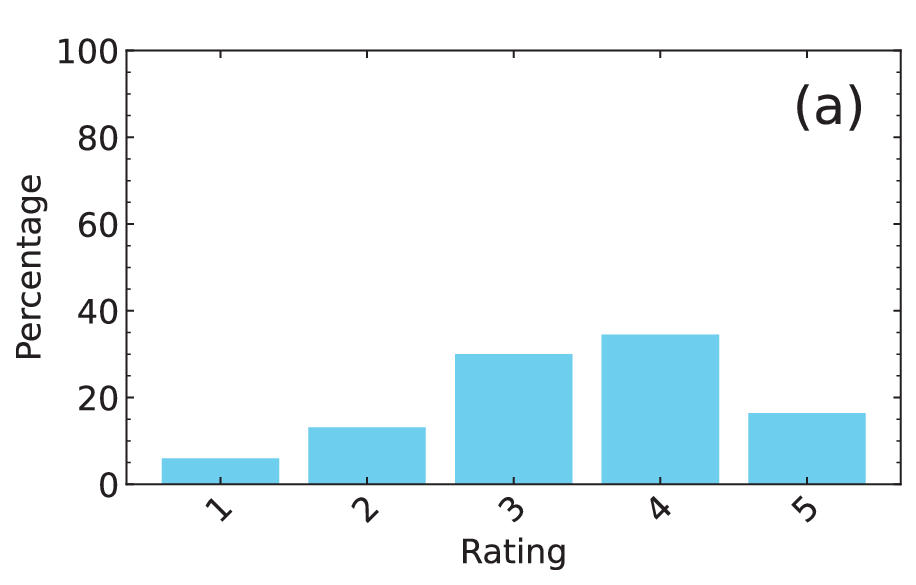}
\includegraphics[width=0.4\linewidth]{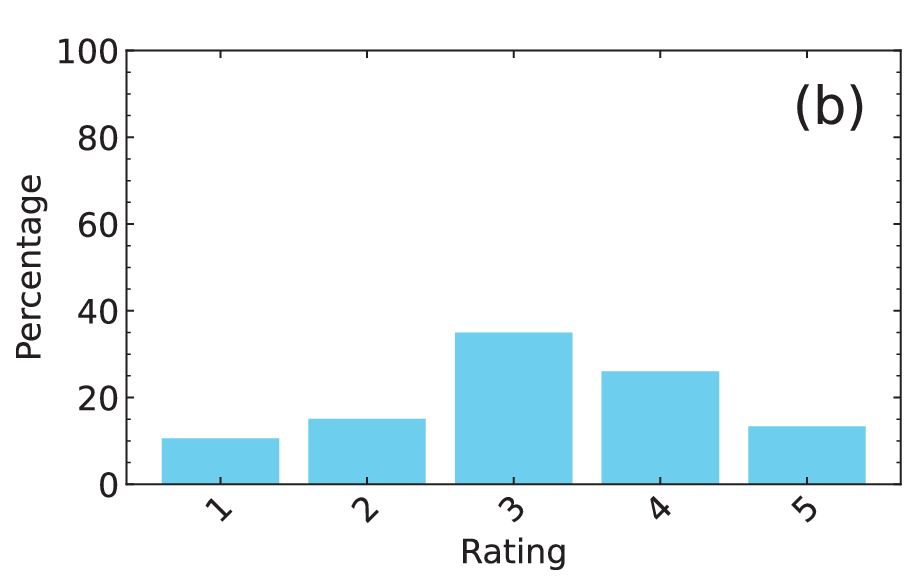}
\includegraphics[width=0.4\linewidth]{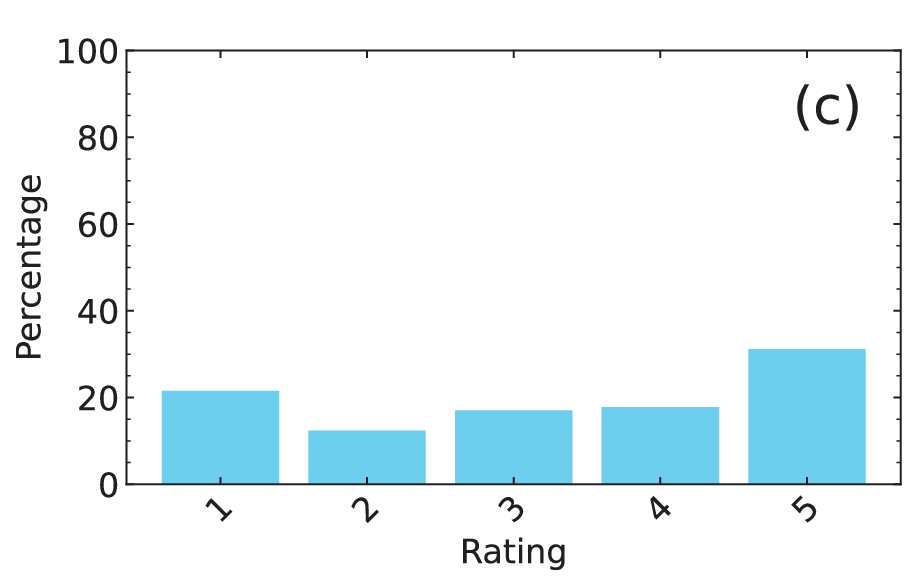}
\includegraphics[width=0.4\linewidth]{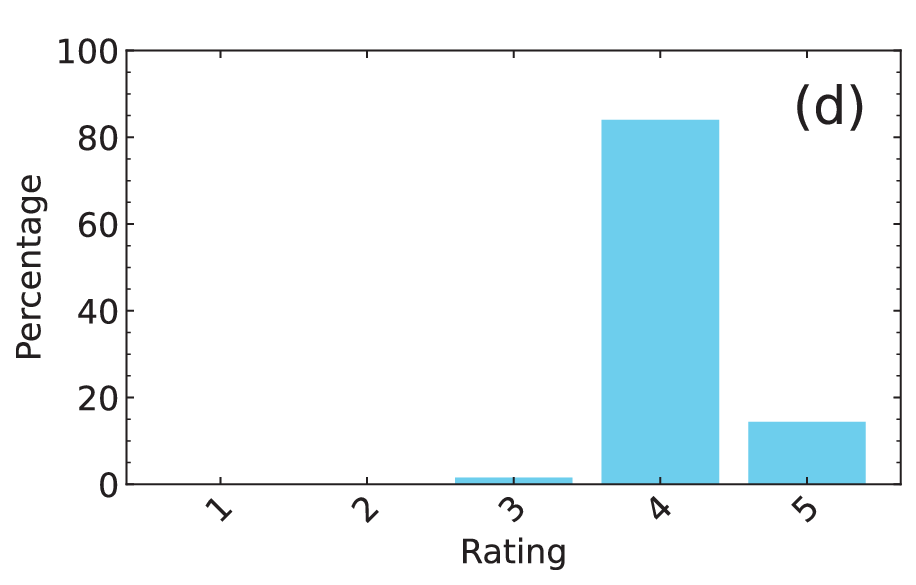}
\caption{
Frequency distributions of ratings expressed as percentages for (a) MovieLens, (b) Netflix, (c) Yahoo, and (d) CiaoDVD datasets.}
\label{fig:frequency}
\end{figure}

Figure~\ref{fig:base} shows a comparison of RMSE and computation time of NBMF and NMF for 10 epochs.
Each epoch involves updating matrix $W$ followed by updating matrix $H$.
The data points represent the averages of RMSE and computation time at each epoch, with some error bars too small to be observed.
Figure~\ref{fig:base} shows that the RMSE decays more rapidly in NBMF than in NMF for all datasets.
Although the difference in the RMSE at each epoch between NBMF and NMF was negligible, the difference among the datasets was remarkable.
The difference among the datasets originates from the frequency distribution of the ratings in each dataset, as shown in Figure~\ref{fig:frequency}.
As elaborated later, when the distribution was sharp and the variance was small, the RMSE tended to be small.
In contrast, when the variance in the frequency was large, the RMSE was relatively large.

\begin{figure}[htbp]
\centering
\includegraphics[width=0.4\linewidth]{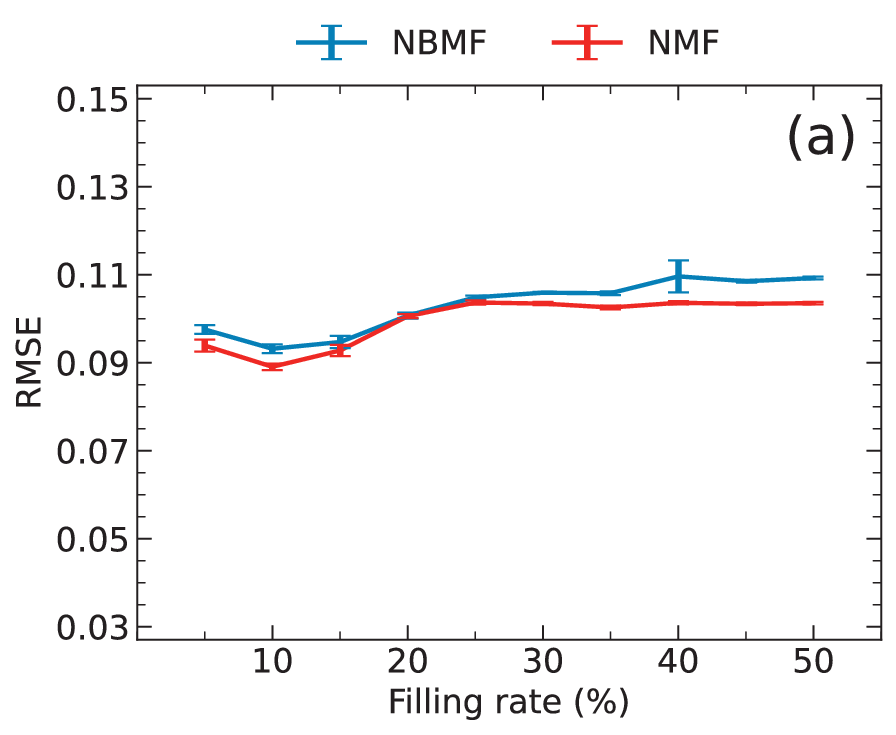}
\includegraphics[width=0.4\linewidth]{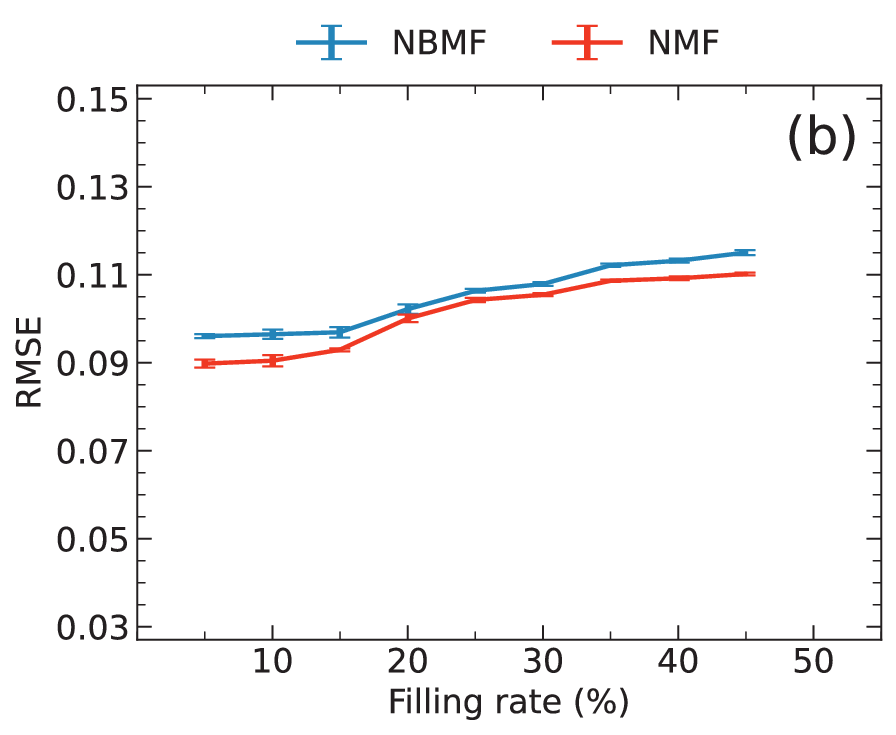}
\includegraphics[width=0.4\linewidth]{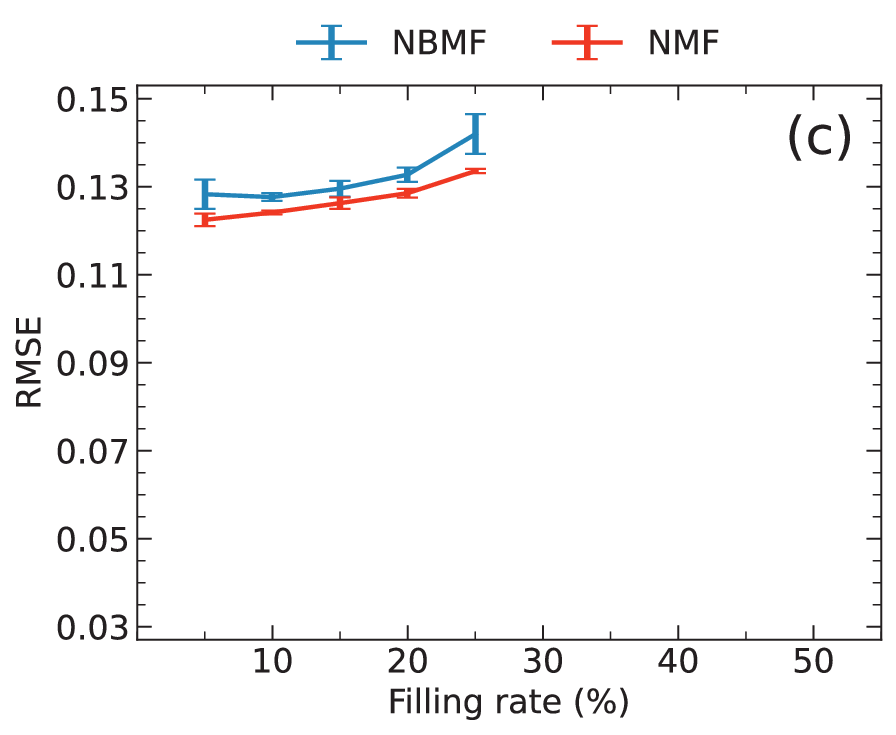}
\includegraphics[width=0.4\linewidth]{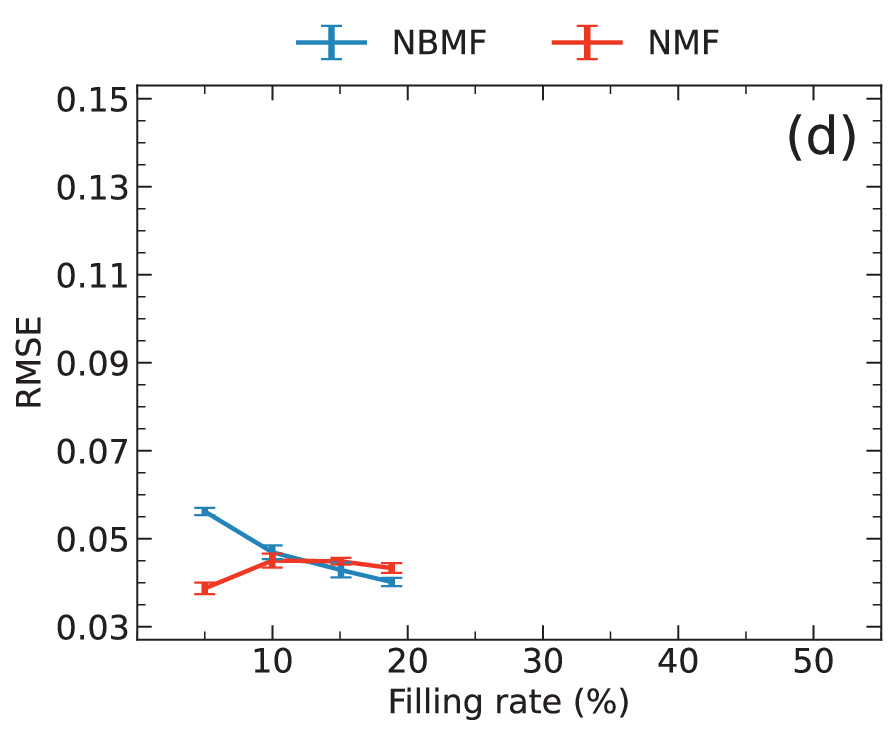}
\caption{Filling rate dependence of RMSE
for (a) MovieLens, (b) Netflix, (c) Yahoo, and (d) CiaoDVD datasets.
Training data with filling rates of more than 45\%, 25\%, and 20\% could not be extracted for (b), (c), and (d), respectively.
The data were averaged over five trials, with error bars denoting the standard deviation.}
\label{fig:filling}
\end{figure}

The filling rate of a rating matrix, which is the proportion of rated elements, influences collaborative filtering.
However, the filling rate dependence of the RMSE varied across the datasets, as shown in Figure~\ref{fig:filling}.
Here, the RMSE was calculated after 10 epochs and averaged over five trials.

NMF is is expected to produce lower RMSE values than NBMF due to its higher resolution.
However, in Figure~\ref{fig:filling}, case (a) demonstrates that the RMSE values for both NBMF and NMF were similar when the filling rate was around 20\%.
Furthermore, in case (d), the RMSE for NBMF was smaller than that for NMF at the filling rate of approximately 20\%.
This inconsistent behavior suggests that the datasets have significant differences in their features.

\begin{figure}[htbp]
\centering
\includegraphics[width=0.4\linewidth]{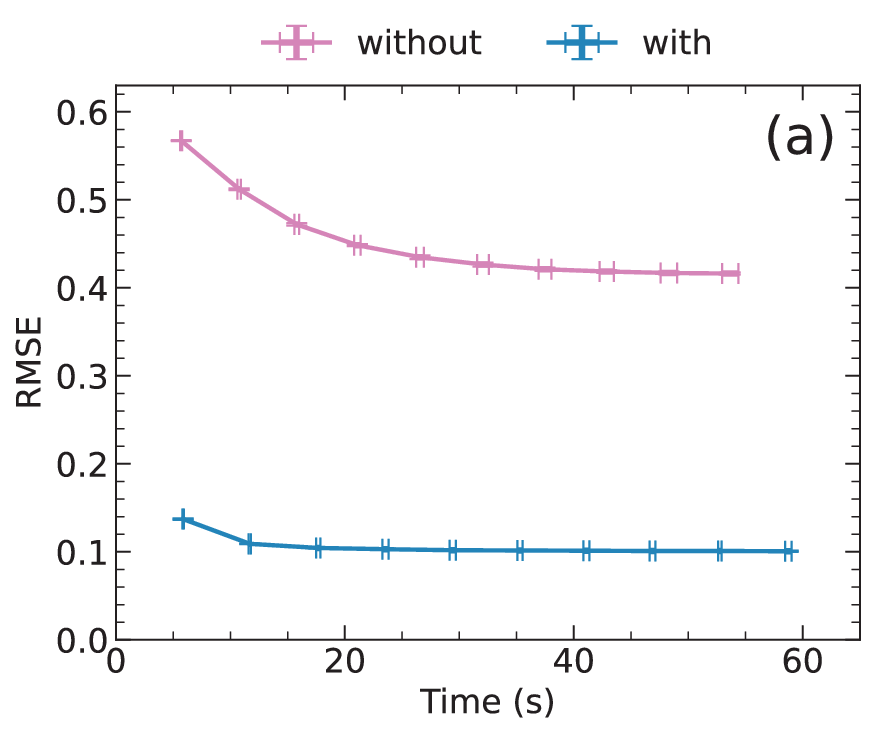}
\includegraphics[width=0.4\linewidth]{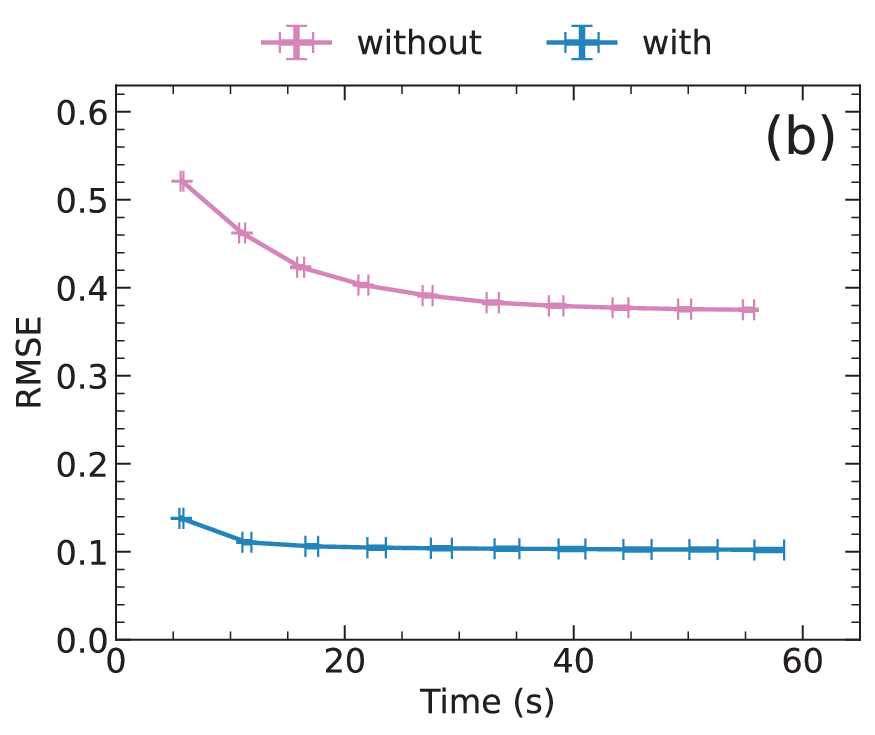}
\includegraphics[width=0.4\linewidth]{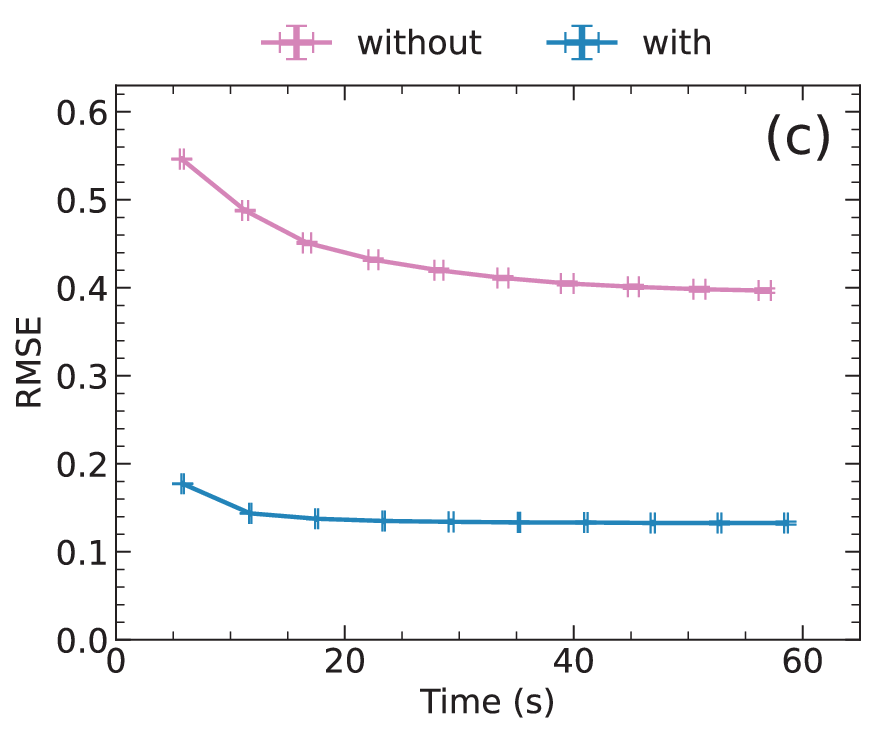}
\includegraphics[width=0.4\linewidth]{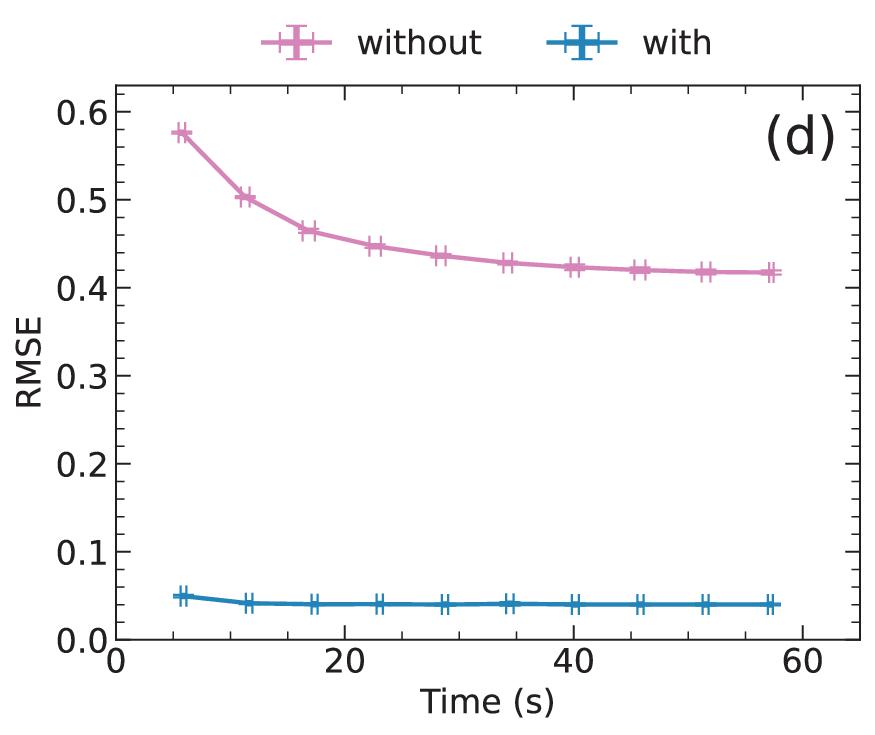}
\caption{RMSE and computation time at each epoch with and without the masking procedure for NBMF for (a) MovieLens, (b) Netflix, (c) Yahoo, and (d) CiaoDVD datasets.
The data were averaged over five trials, with error bars denoting the standard deviation.}
\label{fig:mask}
\end{figure}

Figure~\ref{fig:mask} shows the advantages of this masking procedure.
The masking data for each dataset were the same as those used for NBMF in Figure~\ref{fig:base}.
Notably, the RMSE without the masking procedure was about more than three times larger than the RMSE with the masking procedure for all datasets.
Furthermore, the difference in computation time between the two procedures was minimal.
These findings indicate that the masking procedure provides apparent benefits in collaborative filtering, despite a minor drawback in terms of computation time.

The results indicate that the RMSE reflects specific properties of the data.
Here, we focus on the frequency distribution of ratings, as illustrated in Figure~\ref{fig:frequency}.
The distribution represents the percentage of ratings (1, 2, 3, 4, and 5) among the rated elements in a rating matrix, with a filling rate of 20\%.
The distributions in (a) and (b) showed a broad peak, and the corresponding RMSE had a similar value at 10 epochs in Figures~\ref{fig:base}a and \ref{fig:base}b.
However, the distribution in (c) showed two peaks at 1 and 5, resulting in significant variance.
The corresponding RMSE at 10 epochs in Figure~\ref{fig:base}c was larger than those of  the other three data.
By contrast, the distribution in (d) had a steep peak at 4, indicating that more than 80\% of the rated elements had a value of 4.
The corresponding RMSE at 10 epochs in Figure~\ref{fig:base}d was significantly smaller than those of the other three data.
This observation indicates that a distribution with a sharp peak and  small variance typically results in a smaller RMSE. 
However, a distribution with a broader peak and larger variance often results in a larger RMSE.

\begin{figure}[htbp]
\centering
\includegraphics[width=0.4\linewidth]{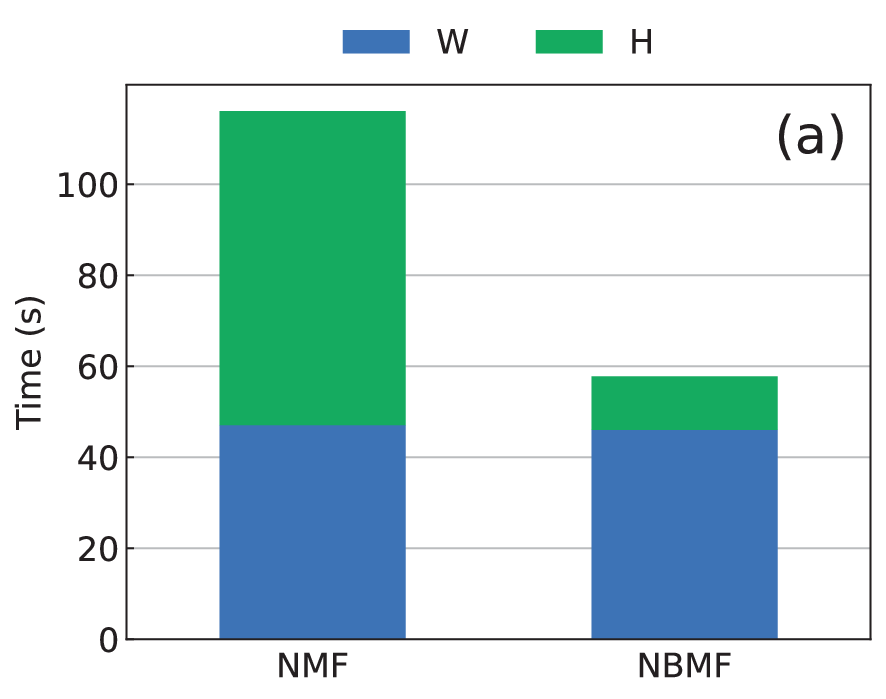}
\includegraphics[width=0.4\linewidth]{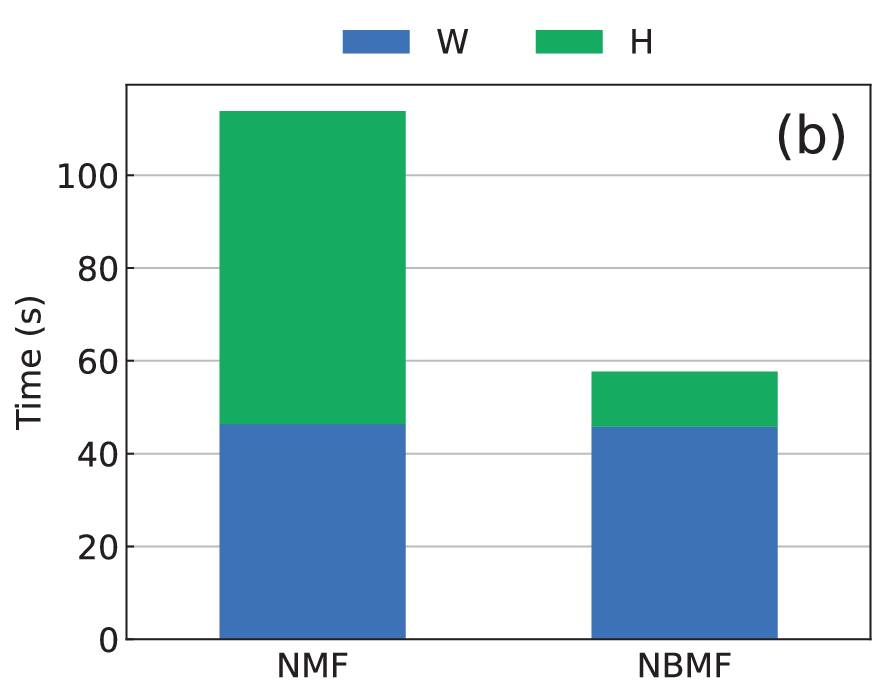}
\includegraphics[width=0.4\linewidth]{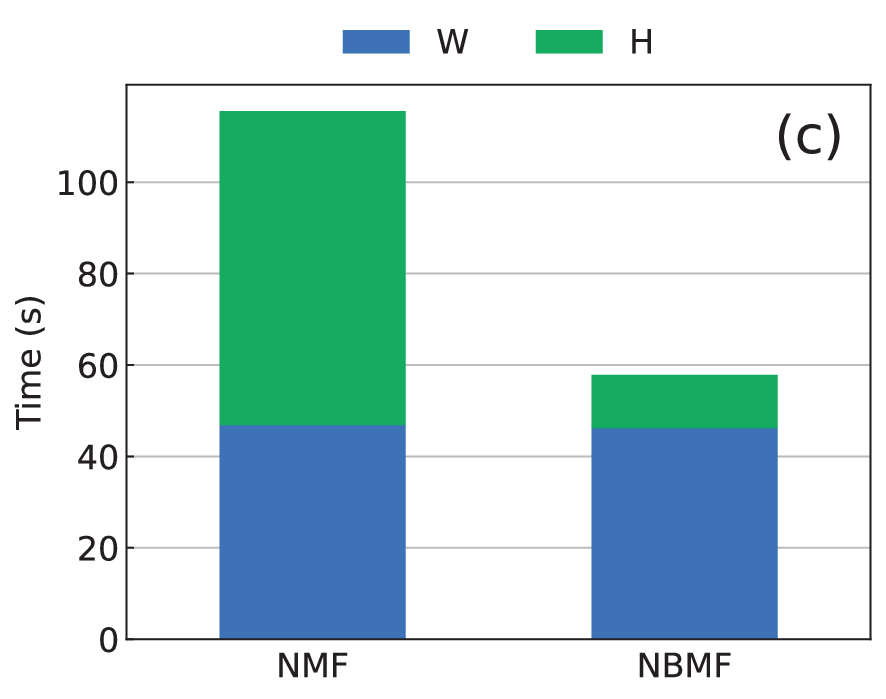}
\includegraphics[width=0.4\linewidth]{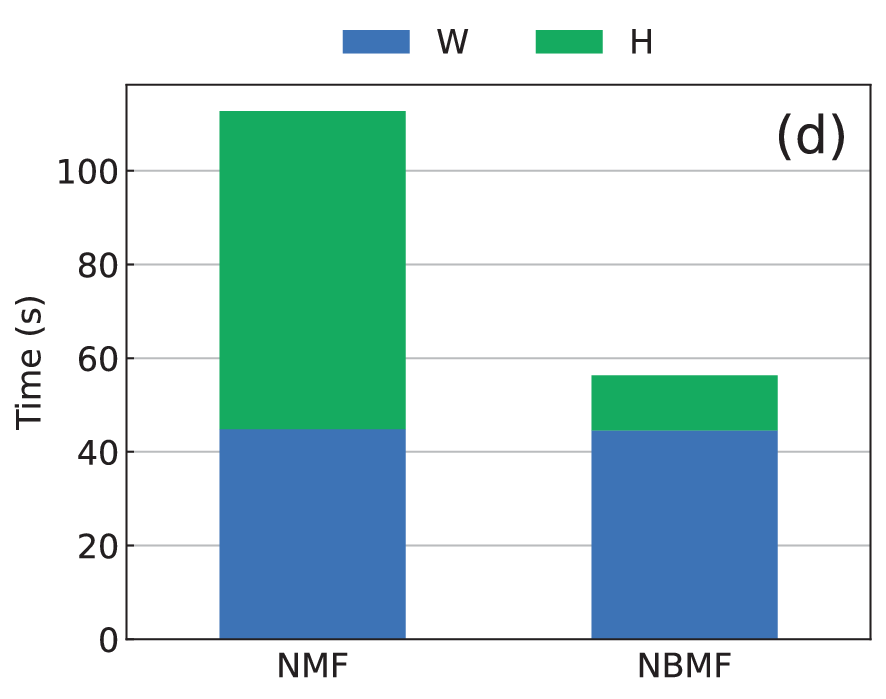}
\caption{Computation time for 10 epochs in NBMF and NMF methods for the (a) MovieLens, (b) Netflix, (c) Yahoo, and (d) CiaoDVD datasets.
The blue and green bars represent the total times spent on updating matrices $W$ and $H$, respectively.}
\label{fig:time}
\end{figure}

The computation time for NBMF was significantly shorter than that for NMF under the same problem setup, as shown in Figure~\ref{fig:time}, for all the datasets.
The total time required to update matrix $W$ over 10 epochs was the same for both NMF and NBMF.
However, the time required to update matrix $H$ in NMF was approximately six times longer than that in NBMF.
This discrepancy suggests that the use of an SB-based machine accelerates the computation to update $H$. 
Additional time is required to minimize Eq.~\eqref{eq:f_H.2} during the update of $H$.
Executing minimization using the SB-based machine involves transforming the objective function into the Ising model form, as explained in Section~\ref{sec:Ising}.
Using the SB-based machine causes the communication time between the CPU and the FPGA, although it is a small fraction of the total time.
Nevertheless, the overall computation time for NBMF, including these additional factors, was shorter than that for NMF.

\begin{figure}[htbp]
\centering
\includegraphics[width=0.4\linewidth]{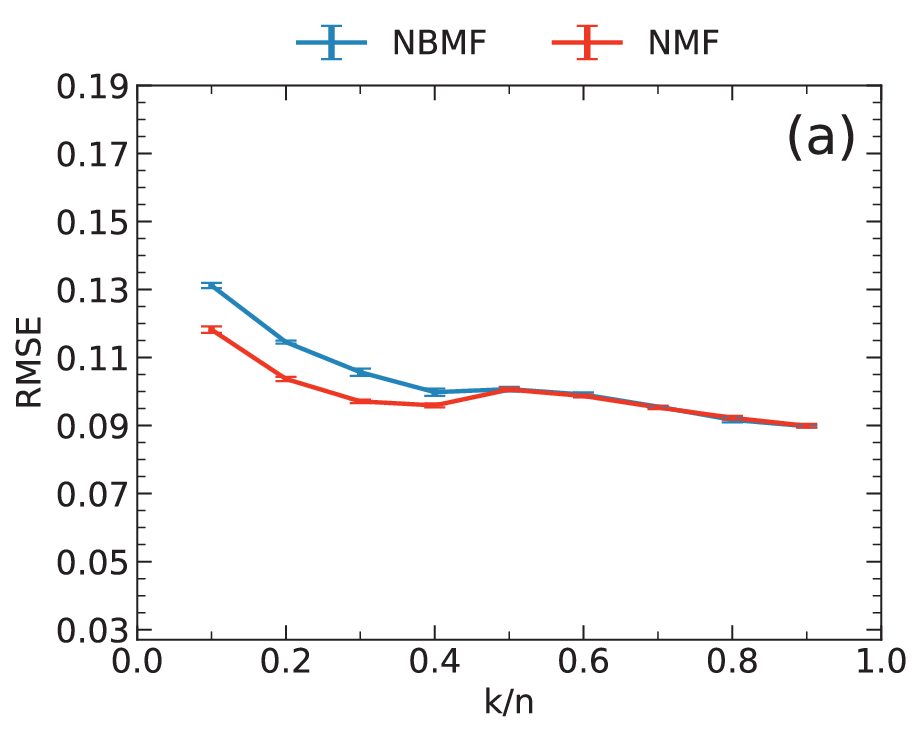}
\includegraphics[width=0.4\linewidth]{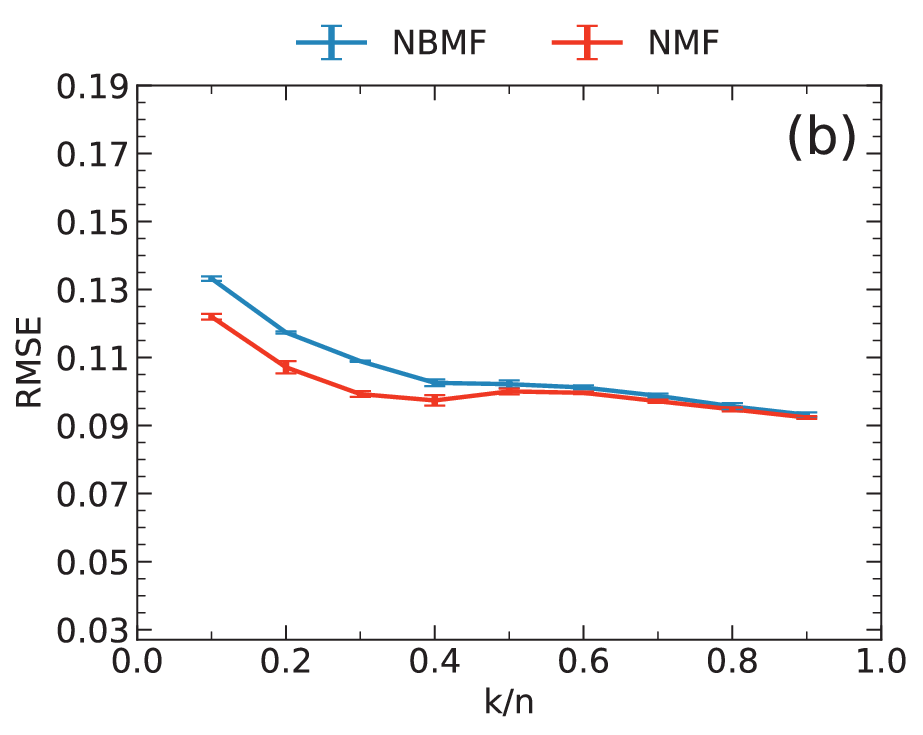}
\includegraphics[width=0.4\linewidth]{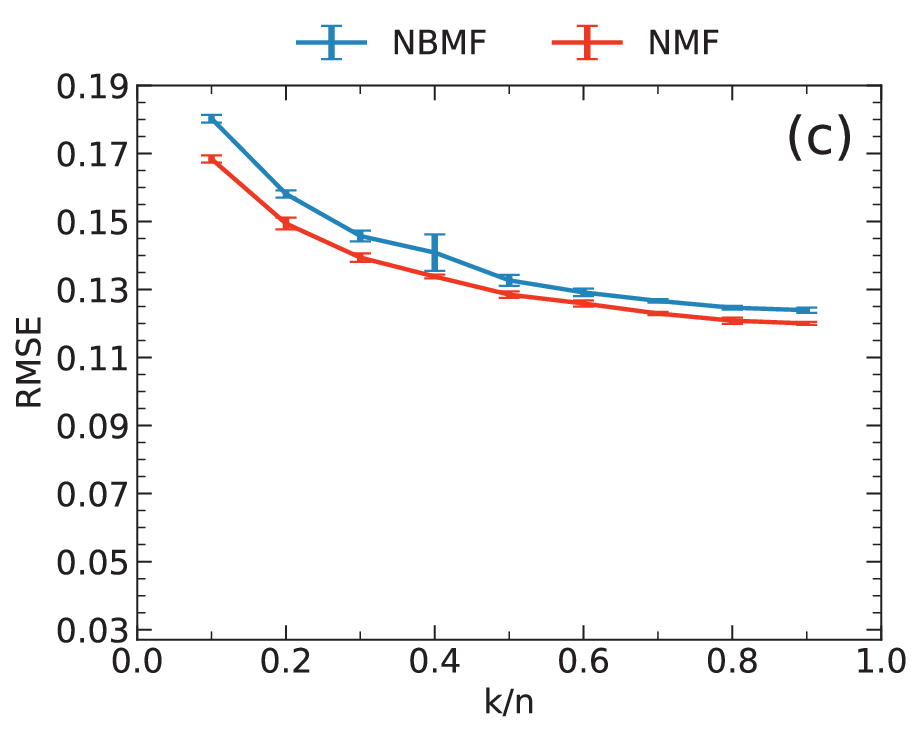}
\includegraphics[width=0.4\linewidth]{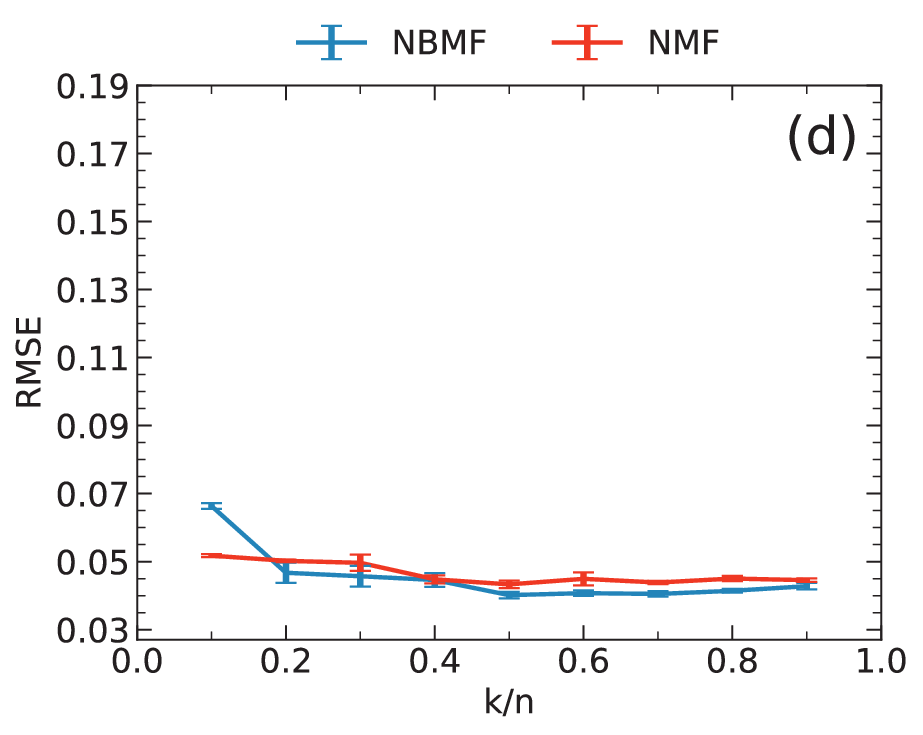}
\caption{RMSE as a function of the ratio of the number of features $k$ to that of users $n$ for (a) MovieLens, (b) Netflix, (c) Yahoo, and (d) CiaoDVD datasets.
The data were averaged over five trials, with error bars denoting the standard deviation.}
\label{fig:feature}
\end{figure}

Throughout this study, the ratio of the number of features to the number of users was fixed at $k/n=0.5$.
In general, the RMSE tends to decrease as the ratio increases.
However, the computation time increases with the ratio because the matrix sizes of $H$ and $W$ increase.
Therefore, a moderate value needs to be selected for this ratio.
As shown in Figure~\ref{fig:feature}, the rate of improvement in the RMSE was slow for ratios of 0.5 or greater across all datasets.
Considering this result, we chose $k/n=0.5$ as an appropriate value for this ratio. 

Our results support the computational advantages of NBMF.
However, several limitations exist. 
Notably, the performance of NBMF is susceptible to the characteristics of datasets.
NMF, which operates on continuous variables, shows comparable or superior accuracy in certain instances compared to NBMF. 
This higher accuracy is due to the greater resolution of continuous representations compared to binary ones.
Furthermore, it is necessary to employ a low-latency system to realize the advantage of computation time.
Even with a high-performance Ising machine, communication overhead between the CPU and the Ising machine can significantly impact overall performance.
Therefore, utilizing a low-latency Ising machine is crucial for efficiently executing NBMF.

\section{Conclusions}
In summary, we proposed a novel approach that employs NBMF with masking for collaborative filtering, and our findings demonstrate a substantial improvement in estimation performance as a result of the masking procedure.
Moreover, our results highlighted the computational advantage of employing an SB-based machine in NBMF.
NBMF with masking can be applied in collaborative filtering across various datasets.
This study reveals the potential of NBMF by utilizing an Ising machine for a wide range of applications.

The efficacy of hybrid concepts using an Ising machine can extend to other algorithms, indicating significant potential for further research in this area. 
Similar hybrid algorithms will become increasingly active in the future as low-latency Ising machines become more advanced and popular.

\section*{Conflict of Interest Statement}

K.T. is an inventor on a Japanese patent application related to this work filed by the Toshiba Corporation (no. P2019-164742, filed 10 September 2019).
This study was conducted as collaborative research between Ochanomizu University and Toshiba Corporation.
The authors declare that they have no other competing interests.

\section*{Author Contributions}

Y.T. and Y.I. conceived and conducted the simulations and analyzed the results.
Y.H. and K.T. discussed improvements in computing and parameter tuning.
K.K. supervised the study and wrote the manuscript.
All the authors reviewed the manuscript.

\section*{Funding}

This study was partially supported by JSPS KAKENHI (Grant Numbers JP23H04499 and JP25H01522) and Murata Science and Education Foundation.

\bibliography{nbmf}

\end{document}